\documentclass[twocolumn,showpacs,preprintnumbers,amsmath,amssymb]{revtex4}
\usepackage{pifont}

\usepackage{graphicx}
\usepackage{dcolumn}
\usepackage{bm}
\usepackage[colorlinks,linkcolor=blue,anchorcolor=blue,citecolor=blue]{hyperref}
\usepackage{lineno} 

\begin{document}
\preprint{}

\title{Continuous time random walks and L\'{e}vy walks with stochastic resetting}

\author{Tian Zhou$^{1}$}
\author{Pengbo Xu$^{1}$}
\author{Weihua Deng$^{1}$}
\affiliation{$^1$School of Mathematics and Statistics, Gansu Key Laboratory of Applied Mathematics and Complex Systems, Lanzhou University, Lanzhou 730000, P.R. China}



\begin{abstract}
Intermittent stochastic processes appear in a wide field, such as chemistry, biology, ecology, and computer science. This paper builds up the theory of intermittent continuous time random walk (CTRW) and L\'{e}vy walk, in which the particles are stochastically reset to a given position with a resetting rate $r$. The mean squared displacements of the CTRW and L\'{e}vy walks with stochastic resetting are calculated, uncovering that the stochastic resetting always makes the CTRW process localized and  L\'{e}vy walk diffuse slower. The asymptotic behaviors of the probability density function of L\'evy walk with stochastic resetting are carefully analyzed under different scales of $x$, and a striking influence of stochastic resetting is observed.

%
%
%
\end{abstract}

\maketitle

\section{Introduction}

Anomalous diffusions are ubiquitous in the natural world, the mean squared displacements (MSD) of which are not linear functions of $t$. The most representative types are  $\big< x^2(t)\big>\sim t^{\alpha}$ with $0<\alpha<1$ or $\alpha>1$, called subdiffusion or superdiffusion. Two popular models for anomalous diffusions are random walks, including continuous time random walk (CTRW) \cite{s11} and the L\'evy walks \cite{s13}, and Langevin equations \cite{coffey}. The models have their own particular advantages under some special backgrounds. For example, by Langevin picture it is relatively easy to model the motion of particles under external potential, while the random walks are ideal models for describing the stochastic resetting.

For the CTRW model, there are two independent random variables (waiting time and jumping length) with given distributions \cite{s16}, which has many applications in finance \cite{s3}, ecology \cite{s4}, and biology \cite{s6}. When the waiting time density has infinite average, such as $\psi(\tau)\sim1/\tau^{1+\alpha}$ with $0<\alpha<1$, and the jump length follows the Gaussian distribution, the CTRW models subdiffusion. If the average of the waiting time is finite and the jump length density $\lambda(x)\sim 1/|x|^{1+\mu}$ with $0<\mu<2$, it is called L\'evy flight, displaying superdiffusion. The other important random walk model is L\'evy walk \cite{s8,s7,s9,s14}, the running time and walking length of which are coupled. The running time $\tau$ is also a random variable analogizing with the waiting time in CTRW model.  The most representative L\'evy walk is the one with constant velocity, and its walking length is then  $v\tau$ in each step.

Intermittent stochastic processes are widely observed in natural world. Considering to search for a target in a crowd, if one can not find it all the time, then the efficient way is to be back to the beginning to start the process again  \cite{evans}.  Besides the stochastic resetting also has a profound use in search strategies in ecology, for example, in intermittent searches for the foraging animals, which consists of long range searching movement and scanning for food and relocation \cite{benichou2005,Lomholt11055,Loverdo2009}. In fact, while trying to find a solution, you may get the impression that you are stuck or got on the wrong track. A natural strategy in such a situation is to reset from time to time and start over \cite{Eule}.  Another way to be generalized is to reset the particle to a dynamic position, such as the maximal location of the particle which is discussed in \cite{Majumdar}, and it is also natural to consider this problem because the intelligent animals who may remember the way they have searched, and they prefer to reset to the maximal frontier that they just traveled. The very early research for reset events is on stochastic multiplicative process \cite{Manrubia1999}, which observes two qualitatively different regimes, corresponding to intermittent and regular behavior. Later in \cite{evans}, the authors build up the master equation for the Brownian walker with a given constant resetting rate $r$ and the resetting position $x_0$; and the work is further generalized to L\'evy flight in \cite{Kusmierz2014}. The stochastic resetting in the Langevin picture is discussed in \cite{Eule}.  Under the framework of CTRW model \cite{Shkilev}, the governing equation of the stochastic resetting problem is established, in which the distribution of the waiting time between the reset events is a sum of an arbitrary number of exponentials. In this paper, we calculate the PDFs of the CTRW and L\'evy walk models with a constant resetting rate, and the used methods are different from \cite{Shkilev}.

This paper is organized as follows. In the second section, we introduce the CTRW with stochastic resetting, get its PDF in Fourier-Laplace space, and do the asymptotic analysis for the PDF and the calculation of the MSD, in which the cases of waiting first and jumping first are respectively discussed. We turn to L\'{e}vy walks with stochastic resetting in the third section; again the model is first introduced, the corresponding PDF in Fourier-Laplace space and the MSDs are calculated, moreover the asymptotic behaviors of PDF when $x$ scales like $t$ and $t^{1/\alpha}$ are also given, which indicates the major effects caused by stochastic resetting. Finally, we conclude the paper with some discussions.

%

\section{Continuous time random walk with stochastic resetting}

In this section we discuss the model of CTRW with stochastic resetting. For the basic theory of CTRW, one can refer to \cite{s11,s12}.
\subsection{Introduction to the models}

Similar to the discussions of ordinary CTRW model, we first consider the `waiting first' process, i.e., the particles will wait for some time $\tau$ with the density $\psi(\tau)$ first, then have a jump $l$ with the density $\lambda(l)$. After finishing a step, the particle will be reset to the given position $x_r$ with the resetting rate $r\in [0,1]$. The resetting only happens at the renewal points. Similarly with the theory of the CTRW model, one can calculate the PDF of the particle staying at position $x$ at time $t$ denoted by $P_w(x,t)$. In order to obtain $P_w(x,t)$, we need to first calculate $Q(x,t)$--the PDF of just having arrived at the position $x$ at time $t$,
\begin{equation}\label{sec2eq5}
\begin{split}
    Q(x,t) =&(1-r)\int_{-\infty}^{\infty}\int_{0}^{t}\lambda(l)\psi(\tau)Q(x-l,t-\tau)d\tau dl \\
      & +r\delta(x-x_r)\int_{-\infty}^{\infty}Q(x,t)dx\\
     & +(1-r)P_0(x,t),
\end{split}
\end{equation}
where $P_0(x,t)$ is the initial condition, being taken as $P_0(x,t)=\delta(x-x_0)\delta(t)$. It can be noted that $x_0$ and $x_r$ may be different. For \eqref{sec2eq5}, the first term on the right hand side represents the particle transiting from position $(x-l)$ at time $(t-\tau)$ to $x$ at time $t$ without resetting with the probability of $1-r$. So is the meaning of the third term.
The second term represents the `resetting', more specifically, the probability that the resetting takes place is $r$ and at the renew moment the resetting can happen at any point, besides the $\delta(x-x_r)$ represents the particle will be reset to position $x_r$ when the resetting happens. Then we begin to consider the PDF $P_w(x,t)$ of the `waiting first' process. There exists
\begin{equation}\label{sec2eq6}
  P_w(x,t)=\int_{0}^{t}\Psi(\tau)Q(x,t-\tau)d\tau,
\end{equation}
where
\begin{equation}\label{survival_prob}
  \Psi(\tau)=\int_{\tau}^{\infty}\psi(\tau')d\tau'
\end{equation}
represents the survival probability. First we consider \eqref{sec2eq5} by taking Fourier transform from $x$ to $k$ defined as $\bar{f}(k)=\mathcal{F}_{x\rightarrow k}\{f(x)\}=\int_{-\infty}^{\infty}e^{-i k x}f(x)dx$ and Laplace transform from $t$ to $s$ defined as $\hat{g}(s)=\mathcal{L}_{t\rightarrow s}\{g(t)\}=\int_{0}^{\infty} e^{-s t}g(t)dt$, which leads to
\begin{equation}\label{sec2eq7}
\begin{split}
 \hat{\bar{Q}}(k,s)=&(1-r)\bar{\lambda}(k)\hat{\psi}(s)\hat{\bar{Q}}(k,s)\\
      &+r e^{-i k x_r}\int_{-\infty}^{\infty}\hat{Q}(x,s)dx+(1-r)e^{-i k x_0}.
 \end{split}
\end{equation}
Now we begin to consider the integral $\int_{-\infty}^{\infty}Q(x,t)dx$. In fact, the stochastic resetting doesn't affect the integral of the density of the particles just arriving at  position $x$ at time $t$, meaning $\int_{-\infty}^{\infty}Q(x,t)dx=\int_{-\infty}^{\infty}q(x,t)dx$, where $q(x,t)$ represents the PDF of the process without resetting just arriving at $x$ at time $t$. When no resetting happens, there exists
\begin{equation}\label{sec2eq13}
  q(x,t)=\int_{-\infty}^{\infty}\int_{0}^{t}\lambda(l)\psi(\tau)q(x-l,t-\tau)dl d\tau + \delta(x-x_0)\delta(t),
\end{equation}
which results in
\begin{equation}\label{sec2eq14}
  \hat{\bar{q}}(k,s)=\frac{e^{-i k x_0}}{1-\lambda(k)\psi(s)}.
\end{equation}
Then
\begin{equation}\label{sec2eq15}
\begin{split}
    \int_{-\infty}^{\infty}\hat{Q}(x,s)dx &=\mathcal{L}_{t\rightarrow s} \Big\{\int_{-\infty}^{\infty} Q(x,t)\Big\}\\
     & =\int_{-\infty}^{\infty}\hat{q}(x,s)dx\\
     &=\hat{\bar{q}}(k=0,s)\\
     &=\frac{1}{1-\hat{\psi}(s)}.
\end{split}
\end{equation}
Substituting \eqref{sec2eq15} into \eqref{sec2eq7} leads to
\begin{equation}\label{sec2eq8}
 \hat{\bar{Q}}(k,s)=\frac{\frac{r e^{-i k x_r}}{1-\hat{\psi}(s)}+(1-r)e^{-i k x_0}}{1-(1-r)\bar{\lambda}(k)\hat{\psi}(s)}.
\end{equation}
Besides, according to \eqref{sec2eq6}, after taking Fourier-Laplace transform w.r.t. $x$ and $t$, respectively, there exists
\begin{equation}\label{sec2eq9}
\begin{split}
   \hat{\bar{P}}_w(k,s) &=\hat{\Psi}(s)\hat{\bar{Q}}(k,s) \\
     & =\frac{1-\hat{\psi}(s)}{s}\hat{\bar{Q}}(k,s)\\
     &=\frac{1}{s}\frac{re^{-i k x_r}+(1-r)(1-\hat{\psi}(s))e^{-i k x_0}}{1-(1-r)\bar{\lambda}(k)\hat{\psi}(s)}.
\end{split}
\end{equation}
It can be checked that $\int_{-\infty}^{\infty}P(x,t)dx=\mathcal{L}^{-1}_{s\rightarrow t}\{P_w(k=0,s)\}=1$, which means the PDF given by \eqref{sec2eq9} is normalized. When $r=0$, then $\hat{\bar{P}}_w(k,s)=\frac{1-\hat{\psi}(s)}{s}\frac{e^{-i k x_0}}{1-\bar{\lambda}(k)\hat{\psi}(s)}$, indicating the recovery of the ordinary CTRW model \cite{s16}; when $r=1$, $\hat{\bar{P}}_w(k,s)=e^{-i k x_r}/s$, i.e., $P_w(x,t)=\delta(x-x_r)$.


Next we begin to consider the `jump first' CTRW model with stochastic resetting, which means that the particle makes a jump at first and then waits for a period of time $\tau$, finally resets to the position $x_r$ with the probability $r$. 
The PDF of finding the `jump first' particle at position $x$ at time $t$, denoted as $P_j(x,t)$, satisfies
\begin{equation}\label{sec2e6}
  P_j(x,t)=\int_{-\infty}^{\infty}dl\int_{0}^{t}\Psi(\tau)\lambda(l)Q(x-l,t-\tau)d\tau.
\end{equation}
Then still by Fourier-Laplace transform w.r.t. $x$ and $t$, respectively, and utilizing \eqref{sec2eq15}, we obtain
\begin{equation}\label{sec2eq16}
\begin{split}
   \hat{\bar{P}}_j(k,s)& =\bar{\lambda}(k)\hat{\Psi}(s)\hat{\bar{Q}}(k,s). \\
     &= \frac{\bar{\lambda}(k)}{s} \frac{re^{-i k x_r}+(1-r)(1-\hat{\psi}(s))e^{-i k x_0}}{1-(1-r)\bar{\lambda}(k)\hat{\psi}(s)}.
\end{split}
\end{equation}
The $P_j(x,t)$ given by \eqref{sec2eq16} still satisfies the normalization condition and recovers the ordinary jump first CTRW model when $r=0$. Besides when $r=1$, it's easy to obtain that $P_j(x,t)=\lambda(x-x_r)$. Comparing with the results of waiting first case, these two kinds of PDFs when the stochastic resetting rate $r=1$ are completely different. The other differences will be shown in the next subsection.

\subsection{Effects of stochastic resetting: the stationary density and MSD}

In this subsection, we mainly discuss the effects of stochastic resetting on the CTRW model. We first consider the stationary density $P^{st}(x)=\lim_{t\rightarrow \infty}P(x,t)$. It is well-known that without an external potential for a general CTRW model, the stationary density is a uniform distribution when the particle moves in a bounded area. According to the final value theorem of Laplace transform, there exists
\begin{equation}\label{fvt}
  \lim_{t\rightarrow \infty}P(x,t)=\lim_{s\rightarrow 0}s \hat{P}(x,s).
\end{equation}
Thus from \eqref{sec2eq9} and \eqref{sec2eq16}, we can obtain
\begin{equation}\label{Pst_waiting}
\begin{split}
   \bar{P}^{st}_w(k) & =\lim_{s\rightarrow 0} \frac{re^{-i k x_r}+(1-r)(1-\hat{\psi}(s))e^{-i k x_0}}{1-(1-r)\bar{\lambda}(k)\hat{\psi}(s)} \\
     & =\frac{re^{-i k x_r}}{1-(1-r)\bar{\lambda}(k)}
\end{split}
\end{equation}
and
\begin{equation}\label{Pst_jump}
\begin{split}
   \bar{P}^{st}_j(k) & =\lim_{s\rightarrow 0} \bar{\lambda}(k)\frac{re^{-i k x_r}+(1-r)(1-\hat{\psi}(s))e^{-i k x_0}}{1-(1-r)\bar{\lambda}(k)\hat{\psi}(s)} \\
     & =\frac{r \bar{\lambda}(k) e^{-i k x_r}}{1-(1-r)\bar{\lambda}(k)}.
\end{split}
\end{equation}
Next we choose the jump length density to be standard Gaussian distribution as an example, i.e., $\lambda(x)=\frac{1}{\sqrt{2\pi}}e^{-x^2/2}$ and substitute it into \eqref{Pst_waiting} and \eqref{Pst_jump}, respectively. Then the stationary density for the waiting first process approximately behaves as
\begin{equation}\label{Pst_wait_appr}
  \bar{P}^{st}_w(k)\sim e^{-i k x_r}\frac{2r}{2r+(1-r)k^2},
\end{equation}
which indicates
\begin{equation}\label{inverse_pst_wait}
  P^{st}_w(x)\sim \sqrt{\frac{r}{2(1-r)}}e^{-\sqrt{\frac{2r}{1-r}}|x-x_r|} ~ {\rm for } ~r\neq 1;
\end{equation}
while for the jump first process, the corresponding stationary probability can be approximately given as
\begin{equation}\label{Pst_jump_appr}
  \bar{P}^{st}_j(k)\sim e^{-i k x_r}\frac{2 r}{2r+k^2},
\end{equation}
and after taking inverse Fourier transform, there is
\begin{equation}\label{inverse_pst_jump}
  P^{st}_j(x)\sim \sqrt{\frac{r}{2}}e^{-\sqrt{2r}|x-x_r|}.
\end{equation}
It should be noted that the stationary results hold under the condition that $x$ is large enough.
\begin{figure}
  \centering
  \includegraphics[width=8cm]{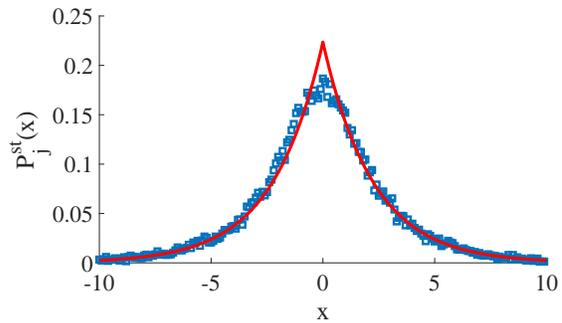}
  \caption{Numerical simulations of the stationary density, $P^{st}_j(x)$, for jump first CTRW process with stochastic resetting, by taking over $10^5$ realizations. Here we choose the waiting time density $\psi(\tau)=\exp(-\tau)$, and the jump length density the standard normal distribution, i.e., $\lambda(x)=\frac{1}{\sqrt{2\pi}}e^{-x^2/2}$, besides the stochastic resetting rate $r=0.1$ and $x_r=0$. The dots are the simulation results at time $t=10^4$, which can be approximately considered as the stationary achieved. The solid curve represents theoretical result. 
  }\label{pst_jumping}
\end{figure}

By \eqref{sec2eq9} and \eqref{sec2eq16}, to calculate the MSD, 
the $\lambda(x)$ and $\psi(\tau)$ need to be specified. Specifically, in this paper we take $\lambda(x)=\frac{1}{\sqrt{2\pi}}e^{-x^2/2}$ with the Fourier transform and the approximate form
\begin{equation}\label{jump_length_appr}
\bar{\lambda}(k)=e^{k^2/2}\sim 1-\frac{k^2}{2}
\end{equation}
and
\begin{equation}\label{power_law}
  \psi(t)=\frac{1}{\tau_0}\frac{\alpha}{(1+\tau/\tau_0)^{1+\alpha}},
\end{equation}
where $\tau_0, \alpha>0$ with the approximation form of Laplace transform,
\begin{equation}\label{waiting_time_appr}
  \hat{\psi}(s)\sim 1-\frac{\tau_0}{\alpha-1}s-\tau_0^\alpha \Gamma(1-\alpha)s^\alpha+\frac{\tau_0^2}{(\alpha-1)(\alpha-2)}s^2
\end{equation}
for $\alpha \neq 1,2$. For the convenience of calculations, in this paper, we choose $\tau_0=1$. In general, the MSD can be obtained by
\begin{equation}\label{MSD}
  \langle x^{2}(t)\rangle=- \frac{d^{2}}{dk^{2}}\bar{P}(k,t)|_{k=0}.
\end{equation}
Substituting \eqref{jump_length_appr} and \eqref{sec2eq9} into \eqref{MSD}, then for waiting first CTRW model, there exists
\begin{equation}\label{MSD_wait_general}
\begin{split}
   \langle x_w^{2}(s)\rangle &= \mathcal{L}_{t\rightarrow s}\{x^2 (t)\}\\
     & =\frac{(1-r)x_0^2+r x_r^2+(1-r)(1-x_0^2)\psi(s)}{s[1-(1-r)\psi(s)]}.
\end{split}
\end{equation}
After some calculations, we obtain
\begin{equation}\label{MSD_wait}
  \langle x_w^{2}(t)\rangle \sim
\begin{cases}
  t^\alpha, & \mbox{if $0<\alpha<1$ and $r=0$},\\
  \frac{1-r+r x_r^2}{r}, & \mbox{if $0<\alpha<1$ and $0<r\leq1$},\\
  t, & \mbox{if $\alpha>1$ and $r=0$},\\
  \frac{1-r+r x_r^2}{r}, & \mbox{if $\alpha>1$ and $0<r\leq1$}.
\end{cases}
\end{equation}
As for the jump first process, we have
\begin{equation}\label{MSD_jump_general}
  \begin{split}
      \langle x_j^{2}(s)\rangle &= \mathcal{L}_{t\rightarrow s}\{x^2 (t)\}\\
       &=\frac{1+(1-r)x_0^2+r x_r^2-(1-r)x_0^2\psi(s)}{s[1-(1-r)\psi(s)]},
  \end{split}
\end{equation}
which implies
\begin{equation}\label{MSD_jump}
  \langle x_j^{2}(t)\rangle \sim
\begin{cases}
  t^\alpha, & \mbox{if $0<\alpha<1$ and $r=0$},\\
  \frac{1+r x_r^2}{r}, & \mbox{if $0<\alpha<1$ and $0<r\leq1$},\\
  t, & \mbox{if $\alpha>1$ and $r=0$},\\
  \frac{1+r x_r^2}{r}, & \mbox{if $\alpha>1$ and $0<r\leq1$}.
\end{cases}
\end{equation}
From the results above, it can be easily concluded that when $0<r<1$, the MSD of the process is a constant, implying the localization, and this constant doesn't depend on the density of waiting time. Besides the jump or waiting first CTRW model will influence the MSD, which is verified by numerical simulations.
\begin{figure}
  \centering
  \includegraphics[width=8cm]{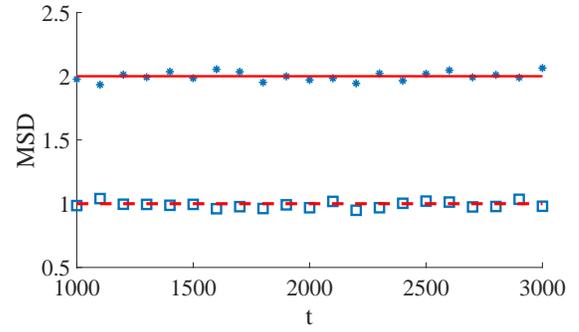}
  \caption{MSDs of jump first and waiting first CTRW model with stochastic resetting. Here we choose the waiting time density as \eqref{power_law} with $\alpha=0.8$, $\tau_0=1$, and the resetting rate $r=0.5$. Both of the start and resetting positions are chosen to be 0, i.e., $x_0=x_r=0$. The stars and squares are obtained by averaging over $10^4$ realizations of jump first and waiting first processes, respectively. The solid and dashed lines are, respectively, the theoretical results shown in \eqref{MSD_jump} and \eqref{MSD_wait}.}\label{pst_jumping}
\end{figure}

\section{L\'{e}vy Walk with Stochastic Resetting}

In this section we begin to discuss the L\'{e}vy walk with stochastic resetting. For the ordinary theory of L\'{e}vy walk, one can refer to \cite{s13}.

\subsection{Introduction to the models}

We still use $Q(x,t)$ to represent the PDF of the particle just arriving position $x$ at time $t$ and having a chance to change the direction, while $P(x,t)$ is the PDF of the particle staying at position $x$ at time $t$. For L\'{e}vy walk, there's no concept of `waiting first' or `jump first'. Here we consider the symmetric L\'{e}vy walk  with a stochastic resetting rate $r$ to the position $x_r$, and the velocity of each step is a constant. The walking time of each step is still denoted as $\tau$ with the density $\psi(\tau)$. The stochastic resetting only happens at the end of each step, i.e., the point which may have chance to change the direction.
%
%
Similarly, there exists
\begin{equation}\label{sec3eq6}
\begin{split}
   Q(x,t)=& (1-r)\int_{-\infty}^{\infty}\int_{0}^{t}\phi(y,\tau)Q(x-y,t-\tau)d\tau dy\\
   &+r\delta(x-x_r)\int_{-\infty}^{\infty}Q(x,t)dx\\
     & +(1-r)P_0(x,t),
\end{split}
\end{equation}
where $P_0(x,t)$ represents the initial condition specified as $P_0(x,t)=\delta(x-x_0)\delta(t)$, and
\begin{equation}\label{sec3eq2}
 \phi(y,\tau)=\frac{1}{2} \delta(|y|-v \tau)\psi(\tau).
\end{equation}
The reason why \eqref{sec3eq6} holds is the same as \eqref{sec2eq5}. 
As for $P(x,t)$, we have
\begin{equation}\label{sec3eq7}
 P(x,t)=\int_{-\infty}^{\infty}\int_{0}^{t}\Phi(y,\tau)Q(x-y,t-\tau)d\tau dy
\end{equation}
and here
\begin{equation}\label{sec3eq4}
 \Phi(y,\tau)=\frac{1}{2}\delta(|y|-v \tau)\Psi(\tau),
\end{equation}
where $\Psi(\tau)$ is the survival probability defined in \eqref{survival_prob}.

The Fourier-Laplace transform of \eqref{sec3eq7} results in
\begin{equation}\label{sec3eq8}
\begin{split}
 \hat{\bar{Q}}(k,s)&=(1-r)\hat{\bar{\phi}}(k,s)\hat{\bar{Q}}(k,s)\\
 & +r e^{-i k x_r}\int_{-\infty}^{\infty}\hat{Q}(x,s)dx+(1-r)e^{-i k x_0},
 \end{split}
\end{equation}
where $\hat{\bar{\phi}}(k,s)=\frac{1}{2}[\hat{\psi}(s+i k v)+\hat{\psi}(s-i k v)]$. For L\'evy walk, the stochastic resetting still doesn't influence the integral $\int_{-\infty}^{\infty}Q(x,t)dx$. According to \cite{s13}, for the L\'evy walk without the stochastic resetting, the density of particle just arriving at position $x$ at time $t$  satisfies
\begin{equation}\label{levy_walk_r=0}
  \hat{\bar{q}}(k,s)=\frac{e^{-i k x_0}}{1-\frac{1}{2}[\hat{\psi}(s+i k v)+\hat{\psi}(s-i k v)]}.
\end{equation}
Then there exists
\begin{equation}\label{q(k=0,s)}
\begin{split}
    \int_{-\infty}^{\infty}\hat{Q}(x,s)dx & =\int_{-\infty}^{\infty}\hat{q}(x,s)dx\\ &=\hat{\bar{q}}(k=0,s)\\
     &=\frac{1}{1-\hat{\psi}(s)}.
\end{split}
\end{equation}
Substituting \eqref{q(k=0,s)} into \eqref{sec3eq8} leads to
\begin{equation}\label{sec3eq9}
 \hat{\bar{Q}}(k,s)=\frac{\frac{r e^{-i k x_r}}{1-\hat{\psi}(s)}+(1-r)e^{-i k x_0}}{1-(1-r)\hat{\bar{\phi}}(k,s)}.
\end{equation}
The Fourier-Laplace transform of (\ref{sec3eq7}) results in
\begin{equation}\label{sec3eq10}
\hat{\bar{P}}(k,s)=\hat{\bar{\Phi}}(k,s)\hat{\bar{Q}}(k,s),
\end{equation}
where $\hat{\bar{\Phi}}(k,s)=\frac{1}{2}[\hat{\Psi}(s+i k v)+\hat{\Psi}(s-i k v)]$, and for survival probability, its Laplace transform satisfies $\hat{\Psi}(s)=\frac{1-\hat{\psi}(s)}{s}$. Finally, the PDF $P(x,t)$ of L\'evy walk with stochastic resetting and the resetting rate $r$ satisfies
\begin{equation}\label{levy walk p(x,t)}
\begin{split}
 &\hat{\bar{P}}(k,s)\\
     &  =\frac{[\hat{\Psi}(s+i k v)+\hat{\Psi}(s-i k v)][\frac{r e^{-i k x_r}}{1-\hat{\psi}(s)}+(1-r)e^{-i k x_0}]}{2-(1-r)[\hat{\psi}(s+i k v)+\hat{\psi}(s-i k v)]},
\end{split}
\end{equation}
the normalization of which can be easily confirmed.

\subsection{Effects of random resetting on L\'evy walks: the stationary density and MSD}

\begin{figure}
  \centering
  \includegraphics[width=8cm]{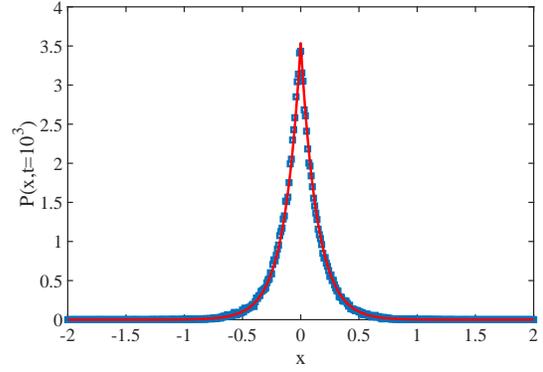}
  \caption{Numerical simulations of stationary density, $P^{st}_{lw}(x)$, for the L\'{e}vy walk with stochastic resetting by sampling over $10^5$ realizations. The velocity $v=0.1$, stochastic rating rate $r=0.5$, the walking time density is exponential distribution $\psi(\tau)=e^{-\tau}$, and the resetting position $x_r=0$. Here the dots represent the simulation results at time $t=10^3$, which is approximately considered as the stationary density. The real line represents the theoretical result.}\label{pst_lw_fig}
\end{figure}

We first discuss the stationary density of L\'evy walk with stochastic resetting and denote it as $P_{lw}^{st}(x)$, which turns out to be significantly different from the stochastic resetting CTRW model.
Here we still use the final value theorem and take $\psi(\tau)=e^{-\tau}$ with the corresponding Laplace transform $\hat{\psi}(s)=\frac{1}{1+s}$. Then according to \eqref{fvt} and \eqref{levy walk p(x,t)}, there is
\begin{equation}\label{pst_levy_walk}
  \bar{P}_{lw}^{st}(k)=e^{-i k x_r}\frac{r}{r+v^2 k^2},
\end{equation}
which indicates
\begin{equation}\label{pst_lw}
  P_{lw}^{st}(x)=\frac{\sqrt{r}}{2v}e^{-\frac{ \sqrt{r}}{v}|x-x_r|},
\end{equation}
being verified by numerical simulations (Fig \ref{pst_lw_fig}).  For the power-law form $\psi(\tau)$ with $0<\alpha<1$ and $1<\alpha<2$ and its Laplace transform \eqref{waiting_time_appr}, the stationary density will be common, i.e., always $0$.

\begin{figure}
  \centering
  \includegraphics[width=8cm]{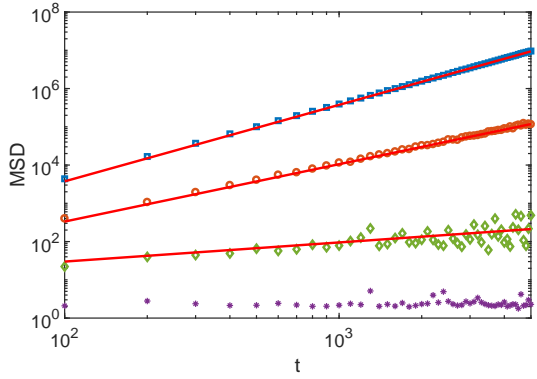}
  \caption{MSDs of stochastic resetting L\'{e}vy walks by sampling over $10^5$ realizations (log-log scale). All the simulations are with the same velocities, the same stochastic resetting rates, and the same resetting positions,  specifically, $v=0.1$, $r=0.5$, and $x_r=0$.
The form of walking density is $\psi(\tau)=\frac{\alpha}{(1+\tau)^{1+\alpha}}$ with $\alpha=0.5$ (squares), $\alpha=1.5$ (circles), $\alpha=2.5$ (diamonds), and $\alpha=3.5$ (dots). The real lines represent the theoretical results. For $\alpha>3$, the simulation results indicate the localization, i.e., MSD is a constant.}\label{msd_lw}
\end{figure}

Further we analyze the MSD $\big<x^2(t)\big>$ of the resetting L\'evy walk with resetting rate $r$ to find out how the stochastic resetting influences the process. First we consider the walking time random variable is exponential distribution, that is, $\psi(\tau)=e^{-\tau}$. Substituting it into \eqref{levy walk p(x,t)} leads to
\begin{equation}\label{sec3eq12}
 \hat{\bar{P}}(k,s)=\frac{\big((1+s ) r e^{-i k x_r}+(1-r) e^{-i k x_0} s  \big)(1+s )}{s(s^2+s +k^{2}v^{2}+s  r+r)}.
\end{equation}
Actually from \eqref{sec3eq12}, we can also obtain \eqref{pst_levy_walk} by taking asymptotic analysis; and the MSD can be got as
\begin{equation}\label{sec3eq13}
\langle x^{2}(s)\rangle \sim \frac{2 v^{2}+r x_r^2}{s(1+s )(r+s)},
\end{equation}
that is
\begin{equation}\label{MSD_exp_LW}
\big<x^2(t)\big>\sim
  \begin{cases}
  2 v^{2} t, & \mbox{if $r=0$},\\
    \frac{2 v^{2}+r x_r^2}{r}, & \mbox{if $0<r\leq 1$}.
  \end{cases}
\end{equation}
It can be seen that for $0<r\leq 1$, the MSD is still a constant. Another representative kind of walking time distribution is $\psi(\tau)=\frac{1}{\tau_0}\frac{\alpha}{(1+\tau/\tau_0)^{1+\alpha}}$, $\tau_0, \alpha>0$, and the corresponding asymptotic form of Laplace transform is \eqref{waiting_time_appr}. For the power-law walking time density, the MSDs of the stochastic resetting L\'evy walk with the resetting rate $r$ are shown in Table \ref{tab1}. One can also conclude from Table \ref{tab1} that $x_0$ and the stochastic resetting position $x_r$ have no influence on the MSD.
\begin{table}
  \centering
  \begin{tabular}{|c|c|c|c|}
    \hline
    ~ & $0<\alpha<1$ & $1<\alpha<2$ & $2<\alpha<3$ \\
    \hline
    $r=0$ &  $(1-\alpha)v^2 t^2$ & $\frac{2 v^2 (\alpha-1)}{(3-\alpha)(2-\alpha)}t^{3-\alpha}$ & $\frac{2 v^2}{\alpha-2} t$ \\
    \hline
    $0<r\leq 1$ &  $\frac{(\alpha-1)(\alpha-2)v^{2}}{2} t^{2}$ & $\frac{(\alpha-1)v^2}{3-\alpha} t^{3-\alpha}$ & $\frac{(\alpha-1)v^2}{3-\alpha} t^{3-\alpha}$ \\
    \hline
  \end{tabular}
  \caption{Asymptotic behavior of MSD for the stochastic resetting L\'evy walk, the walking time density of which is power-law with the asymptotic behavior in Laplace space shown in \eqref{waiting_time_appr} and $\alpha$  chosen in different regions.}\label{tab1}
\end{table}
From the above analysis, it can be seen that the differences between the MSDs of the stochastic resetting CTRW and L\'evy walk are apparent.
Specifically, the former one is always localized when the stochastic resetting rate $0<r\leq 1$ while for the latter one there is no such localization when the walking density is power-law and $0<\alpha<3$. The numerical simulations shown in Fig. \ref{msd_lw} verify the results, besides Fig. \ref{msd_lw} also indicates that the localization will emerge when $\alpha>3$.

Next we consider the infinite density \cite{s14} of stochastic resetting L\'evy walk. It is an effective technique to capture some of the important statistical messages of the stochastic process.
In the following, we would consider two different scales of $x$, and the influence of stochastic resetting on the PDF will be shown.

\subsection{Infinite density of the stochastic resetting L\'evy walk when $x$ and $t$ are of the same scale}

Here we focus on $x_r=x_0=0$ and the power-law walking time density with $1<\alpha<2$. Besides for the convenience of calculations, we take
\begin{equation}\label{psi_1<alpha<2}
\hat{\psi}(s)\sim 1-\big<\tau\big>s+A s^\alpha,
\end{equation}
where according to \eqref{waiting_time_appr} we have $\big<\tau\big>=\tau_0/(\alpha-1)$ and $A=-\tau_0^\alpha \Gamma(1-\alpha)$ with $\tau_0=1$.
First we consider $x$ and $t$ of the same scale, that is, the ratio $x/t$ or $k/s$ is a constant. Substituting \eqref{psi_1<alpha<2} into \eqref{levy walk p(x,t)} leads to
\begin{widetext}
\begin{equation}\label{sec4eq4}
  \hat{\bar{P}}(k,s)\sim \frac{1}{s}\left(\frac{r}{\big<\tau\big> s-A s^{\alpha}}+1-r\right) \frac{2\big<\tau\big>-A s^{\alpha-1}[(1+\frac{i k v}{s})^{\alpha-1}+(1-\frac{i k v}{s})^{\alpha-1}]}{\frac{2r}{k}\frac{k}{s}+(1-r)
  \left\{2\big<\tau\big>-A s^{\alpha-1}[(1+\frac{i k v}{s})^{\alpha}+(1-\frac{i k v}{s})^{\alpha}]\right\}}.
\end{equation}
\end{widetext}
For $r=0$, the PDF shown in \eqref{sec4eq4} reduces to the case discussed in \cite{s14}. 

When $0<r\leq 1$, there is the asymptotic form
 \begin{equation}\label{sec4eq5}
   \hat{\bar{P}}(k,s)\sim \frac{1}{s}\bigg(1-\frac{A s^{\alpha-1}[(1+\frac{i k v}{s})^{\alpha-1}+(1-\frac{i k v}{s})^{\alpha-1}]}{2 \big<\tau\big>}\bigg).
\end{equation}
Note that
\begin{equation}\label{sec4eq6}
\begin{split}
    &(1+\xi)^{\alpha-1}+(1-\xi)^{\alpha-1}\\
    &~~~~=2+2\sum_{m=1}^{\infty}
 \frac{(\alpha-1)\cdots(\alpha-2m)}{(2m)!}\xi^{2m} \\
     & ~~~~=2+2\sum_{m=1}^{\infty}\frac{(-\alpha)_{2m}(\alpha-2 m) \xi^{2 m}}{\alpha (2 m)!},
\end{split}
\end{equation}
 where $(\alpha)_m=\Gamma(\alpha+m)/\Gamma(\alpha)=\alpha(\alpha+1)\cdots(\alpha+m-1)$ is the Pochhammer symbol.
Substituting \eqref{sec4eq6} into \eqref{sec4eq5} results in
\begin{equation}\label{sec4eq7}
 \begin{split}
 & \hat{\bar{P}}(k,s) \sim\frac{1}{s}-\frac{A s^{\alpha-2}}{\big<\tau\big>}\\
 & ~~~~  -\sum_{m=1}^{\infty}\frac{A (-\alpha)_{2m}(\alpha-2m)(-1)^{m}v^{2m}k^{2m}}{\alpha \langle\tau\rangle (2 m)!}s^{\alpha-2-2 m}.
  \end{split}
\end{equation}
Then according to the inverse Laplace transform $\mathcal{L}^{-1}_{s\rightarrow t}\{s^{\alpha-2-2 m}\}=t^{2 m-\alpha+1}/\Gamma(2 m-\alpha+2)$, after the long-time limit we obtain the asymptotic form
\begin{equation}\label{sec4eq8}
 \begin{split}
  \bar{P}_A(k,t) &\sim 1-\frac{B t^{1-\alpha}}{1-\alpha}-B t^{1-\alpha}\sum_{m=1}^{\infty}\frac{(-1)^m (k v t)^{2 m}}{(2 m)!(2 m-\alpha+1)}\\
  & \sim 1-\frac{Bt^{1-\alpha}}{1-\alpha}-Bt^{1-\alpha}\tilde{B}_{\alpha-1}(k v t),
\end{split}
\end{equation}
where $B=\frac{A}{\Gamma(1-\alpha)\langle\tau\rangle}<0$ for $1<\alpha<2$ and $\tilde{B}_{\alpha}(y)=\sum_{m=1}^{\infty}\frac{(-1)^m y^{2 m}}{(2 m)!(2 m-\alpha)}$.
With the expansion,
$\cos(y)=\sum_{n=0}^{\infty}\frac{(-1)^n y^{2 n}}{(2 n)!}$, it is also easy to see $\tilde{B}_{\alpha}(y)=\int_{0}^{1}\frac{\cos(wy)-1}{w^{1+\alpha}}dw$.
Then applying the inverse Fourier transform, defined as
\begin{equation}\label{inverse_fourier}
  f(x)=\mathcal{F}^{-1}_{k\rightarrow x}\{\bar{f}(k)\}=\frac{1}{2\pi}\int_{-\infty}^{\infty}
  e^{i k x}\bar{f}(k)dk,
\end{equation}
on $\tilde{B}_{\alpha}(k v t)$ and denoting the corresponding inverse transform function by $\tilde{B}_{\alpha}(x, v t)$, there exists
\begin{equation}\label{sec4eq9}
\begin{split}
\tilde{B}_{\alpha}(x,vt)&=\frac{1}{2\pi}\int_{-\infty}^{\infty}e^{i k x}\tilde{B}_{\alpha}(k v t)dk\\
     & =\frac{1}{2\pi}\int_{-\infty}^{\infty}e^{i k x}dk\int_{0}^{1}\frac{\cos(wkvt)-1}{w^{1+\alpha}}dw\\
     &=
     \begin{cases}
       \frac{1}{2}\frac{(vt)^{\alpha}}{|x|^{1+\alpha}}, & \mbox{if}~~ |x|<v t \\
       0, & \mbox{if}~~ |x|>v t,
     \end{cases}
     \end{split}
\end{equation}
where $\mathcal{F}^{-1}_{k\rightarrow x}\{\cos(ky)\}=\frac{1}{2}[\delta(x-y)+\delta(x+y)]$ is used. It's easy to conclude that $\tilde{B}_{\alpha}$ isn't integrable w.r.t. $x$ though the interval $(-\infty,\infty)$. Finally, from the inverse Fourier transform of \eqref{sec4eq7} and \eqref{sec4eq9}, we can obtain
\begin{equation}\label{sec4eq10}
\begin{split}
   P(x,t)&\sim \delta(x)-\frac{B t^{1-\alpha}}{1-\alpha}\delta(x)-Bt^{1-\alpha}\tilde{B}_{\alpha-1}(x,vt)\\
     & =\delta(x)-\frac{B t^{1-\alpha}}{1-\alpha}\delta(x)-\frac{B}{2 }\frac{v^{\alpha-1}}{|x|^{\alpha}}.
\end{split}
\end{equation}
Thus when the stochastic resetting rate $0<r\leq 1$ and $x\neq 0$, the distribution of $x$ near the points $\pm v t$ has the form
\begin{equation}\label{infinite_density}
  P(x,t)\sim \frac{(\alpha-1)v^{\alpha-1}}{2|x|^\alpha},
\end{equation}
which is verified by the simulations (see Fig. \ref{id}).
It can be noted that although the form of (\ref{infinite_density}) doesn't explicitly depend on $t$, the region where the approximation makes sense depends on $t$. Besides the stochastic resetting doesn't influence the infinite density. 



\subsection{The density when $x$ and $t^{1/\alpha}$ are of the same scale}

In the previous subsection we focus on the approximation behaviour around the points $\pm vt$, i.e., the scales of $x$ and $t$ are the same. Now, we turn to consider $x \propto t^{\frac{1}{\alpha}}$, that is, the ratio $k^\alpha/s$ always is a constant, which means the approximation of the central part of PDF denoted as $P_{cen}(x,t)$. Besides in this subsection, the calculations are under the condition of $|s|\ll|k v|\ll1$.


When $r=0$ and $x \propto t^{\frac{1}{\alpha}}$, Eq. \eqref{levy walk p(x,t)} reduces to the PDF of ordinary L\'evy walk discussed in \cite{s14}. Here we focus on the differences when stochastic resetting is involved.
For $r=0$, 
\begin{equation}\label{sec4eq16}
  \hat{\bar{P}}_{cen}(k,s) \sim \frac{1}{s+|k|^{\alpha}K_{\alpha}},
\end{equation}
where $K_{\alpha}=-\cos\frac{\alpha\pi}{2} \frac{A}{\langle\tau\rangle} |v|^{\alpha}, 1<\alpha<2$, is the anomalous diffusion coefficient. According to \cite{s17,s15,s16,s7,s21}, one can take the inverse Laplace transform of (\ref{sec4eq16}) w.r.t. $s$ to get
\begin{equation}\label{sec4eq17}
 \bar{P}_{cen}(k,t)\sim e^{-K_{\alpha}t|k|^{\alpha}}.
\end{equation}
With the help of symmetrical L\'{e}vy density $L_{\alpha}(y)$ \cite{s17,s16,s22}, the inverse Fourier transform of \eqref{sec4eq17} yields a symmetric L\'{e}vy stable PDF
\begin{equation}\label{sec4eq18}
 P_{cen}(x,t)\sim \frac{1}{(K_{\alpha}t)^{\frac{1}{\alpha}}} L_{\alpha}\left[\frac{x}{(K_{\alpha}t)^{\frac{1}{\alpha}}}\right].
\end{equation}
For large $|x|$, there is an asymptotic behavior of $L_{\alpha}(x)\propto |x|^{-(1+\alpha)}$.

\begin{figure}
  \centering
  \includegraphics[width=8cm]{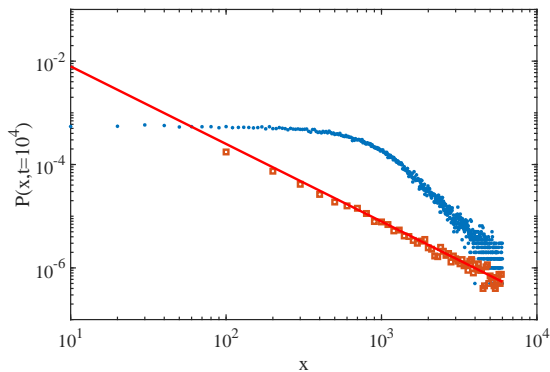}
  \caption{Numerical simulations of the density of L\'{e}vy walks with or without stochastic resetting at time $t=10^4$ by sampling over $2\times 10^5$ realizations (log-log scale). The walking time density $\psi(\tau)=\frac{\alpha}{(1+\tau)^{1+\alpha}}$ with $\alpha=1.5$. The velocity $v=1$, the resetting position $x_r=0$. The circles and squares are the simulation results with stochastic resetting rate $r=0$ and $r=0.1$, respectively. While the real line represents the theoretical result for the latter case. It turns out that there is no change on  the form of the density as the scale of $x$ is different when $r\neq 0$.}\label{id}
\end{figure}

For the case of $0<r\leq 1$, according to \eqref{levy walk p(x,t)}, we conclude
\begin{equation}\label{sec4eq19}
\begin{split}
    & \hat{\bar{P}}_{cen}(k,s) \\
     &\sim \frac{r}{\big<\tau\big>s} \frac{2\big<\tau\big>-A[(s+ikv)^{\alpha-1}+ (s-ikv)^{\alpha-1}]}{2r+(1-r)\{2\big<\tau\big>s-A [(s+ikv)^{\alpha}+ (s-ikv)^{\alpha}]\}}\\
     &\sim \frac{r}{\big<\tau\big>s} \frac{2\big<\tau\big>-A[(ikv)^{\alpha-1}+ (-ikv)^{\alpha-1}]}{2r+(1-r)\{2\big<\tau\big>s-A [(ikv)^{\alpha}+(ikv)^{\alpha}]\}} \\
     &\sim \frac{2\big<\tau\big>-A[(ikv)^{\alpha-1}+ (-ikv)^{\alpha-1}]}{2\big<\tau\big>s}.
\end{split}
\end{equation}
Note that
\begin{equation}\label{sec4eq20}
  (ikv)^{\alpha-1}+ (-ikv)^{\alpha-1}=2 |k|^{\alpha-1} v^{\alpha-1} \cos\left(\frac{ \alpha-1}{2} \pi\right).
\end{equation}
Substituting \eqref{sec4eq20} into \eqref{sec4eq19} leads to
\begin{equation}\label{sec4eq21}
  \hat{\bar{P}}_{cen}(k,s)\sim \frac{\big<\tau\big>-A \cos\big(\frac{ \alpha-1}{2} \pi\big) v^{\alpha-1} |k|^{\alpha-1} }{\big<\tau\big> s}.
\end{equation}
Taking inverse Fourier transform $k\rightarrow x$ results in
\begin{equation}\label{sec4eq22}
\begin{split}
  & \hat{P}_{cen}(x,s) \\
     & \sim\frac{\delta(x)}{s}-\frac{A v^{\alpha-1}}{\pi\big<\tau\big> s}\cos\left(\frac{\alpha-1}{2}\pi\right)\cos \left(\frac{\alpha\pi}{2}\right)|x|^{-\alpha} \Gamma(\alpha).
\end{split}
\end{equation}
Further taking inverse Laplace transform $s\rightarrow t$ and considering the definitions of $A$ and $\big<\tau\big>$, there is
\begin{equation}\label{sec4eq23}
\begin{split}
   &P_{cen}(x,t)\\
    & \sim \delta(x)-\frac{1}{\pi}\Gamma(2-\alpha)\Gamma(\alpha) v^{\alpha-1}\cos\big(\frac{\alpha-1}{2}\pi\big) \cos\big(\frac{\alpha\pi}{2}\big)|x|^{-\alpha} \\
     & =\delta(x)+\frac{1}{2}(\alpha-1)v^{\alpha-1}|x|^{-\alpha}.
\end{split}
\end{equation}
When $0<r\leq 1$ and $x \neq 0$, there holds
\begin{equation}\label{sec4eq24}
  P_{cen}(x,t)\sim \frac{1}{2}(\alpha-1)v^{\alpha-1}|x|^{-\alpha},
\end{equation}
which is the same as the result of \eqref{infinite_density} obtained by the same scales of $x$ and $t$. This is a striking difference between the ordinary L\'evy walk and the stochastic resetting one. Specifically, the asymptotic behaviors of the PDF of the former process are different when $x$ scales like $t$ and $t^{1/\alpha}$, while the ones of the latter one are the same; see Fig. \ref{id}. 


\section{Conclusion}

The phenomena of stochastic resetting are often observed in the natural world. This paper focuses on the CTRW model and L\'evy walk with stochastic resetting. A series of theoretical results are established. The striking observations include: 1) both the MSDs of the waiting first and jump first CTRW models are constant (though different), which implies localization; 2) when $t$ is large enough and $r\neq 0$, the distribution of waiting time does not influence the MSDs of CTRWs; 3) the different walking time distributions can affect the MSDs of L\'evy walks; 4) when the walking time density of L\'evy walk is power law with $1<\alpha<2$, the asymptotic behaviors are different when $x$ scales like $t$ and $t^\alpha$, while for $0<r\leq 1$ the asymptotic behaviors are the same for both cases. The possible further research projects can be: to consider the processes with stochastic resetting not at the renewal points; to discuss the process with variable resetting positions, e.g., the maximum position.

\section*{Acknowledgements}
This work was supported by the National Natural Science Foundation of China under grant no. 11671182, and the Fundamental Research Funds for the Central Universities under grant no. lzujbky-2018-ot03.

\bibliography{ref}

\end{document}